\documentclass[%
 reprint,
 amsmath,amssymb,
 aps,
]{revtex4-2}
\usepackage{physics}
\usepackage{graphicx}% Include figure files
\usepackage{dcolumn}% Align table columns on decimal point
\usepackage{bm}% bold math
\usepackage{dsfont}
\usepackage{subcaption} % for subfigure environment
\usepackage{caption}
\usepackage{xcolor}
\usepackage{braket}

\usepackage{algorithm}
\usepackage{algpseudocode}

\usepackage{hyperref}
\usepackage{mathtools}

\begin{document}

\preprint{APS/123-QED}

\title{Pauli Spectrum and Stabilizer R\'enyi Entropy in Gapless Symmetry-Protected Topological Phases}% Force line breaks with \\

\author{Ying-Lin Li}
 \email{s1012424@gmail.com}
\author{Po-Yao Chang}%
 \email{pychang@phys.nthu.edu.tw}
\affiliation{%
Department of Physics, National Tsing Hua University, Hsinchu 30013, Taiwan
}%

\date{\today}% It is always \today, today,

\begin{abstract}
Quantum entanglement is widely used as a diagnostic of topological phases of matter. Beyond entanglement, non-stabilizerness captures a distinct aspect of quantum many-body states by quantifying their distance from the manifold of stabilizer states. In this work, we study the stabilizer R\'enyi entropy in symmetry protected topological (SPT) phases, including both gapped SPT, non-intrinsically gapless SPT, and intrinsically gapless SPT phases. Under symmetry preserving perturbations, we find numerically that the stabilizer R\'enyi entropy exhibits an extremum near the phase transition. However, the stabilizer R\'enyi entropy alone cannot distinguish different SPT phases. In contrast, the Pauli spectrum reveals a characteristic crossing structure at the transition point. This crossing reflects the exchange of dominant Pauli-string correlations associated with the non-local string order parameters of the two topological distinct phases. 
For gapped SPT and non-intrinsically gapless SPT phases, the crossing structure can be understood from a local-unitary duality that maps the Pauli spectrum between the two phases. For intrinsically gapless SPT phases, such a local-unitary mapping is absent. Instead, we find that the Pauli-spectrum mapping is generated by a non-invertible duality transformation. These results show that although the stabilizer R\'enyi entropy provides only a coarse diagnostic of phase transitions, the Pauli spectrum contains finer information about the exchange of string-order sectors. Our findings demonstrate that quantum magic offers a complementary perspective for characterizing both gapped and gapless SPT phases.
\end{abstract}

\maketitle

%\tableofcontents

\section{\label{sec:Introduction}Introduction}
Symmetry-protected topological (SPT) phases are paradigmatic examples of quantum matter whose characterization relies on symmetry rather than local order parameters. While gapped SPT phases are well understood~\cite{Chen_2011, Schuch_2011, Pollmann_2012, Chen_2013, Senthil_2015}, recent work has extended these ideas to gapless systems, leading to the notion of gapless SPT or more broadly symmetry-enriched criticality~\cite{achain_Verresen_2017, Verresen_2018, Jones_2019, Verresen_2021, Thorngren_2021, YU20261}. In this setting, symmetries can distinguish phases or critical theories that otherwise belong to the same universality class. These ideas have led to many subsequent developments and realizations~\cite{Li_2024, Li_2025, wen_2023_topoholo, Wen_2023_intrinsic, Yu_2024, Cheng_2011, Keselman_2015, Scaffidi_2017, Iemini_2015, Parker_2018, Yu_2022}. %{\color{blue}Accordingly, diagnostics that explicitly incorporate the protecting symmetry can provide information beyond conventional universal data. {\color{red}[PYC: add references?] }}

A key class of diagnostics for SPT phases is provided by non-local string order parameters~\cite{Nijs_1989, Kennedy_1992, Garca_2008, Pollmann_2012_SPT, Else_2013}, which encode symmetry fractionalization~\cite{Chen_2011, Schuch_2011, Pollmann_2012, achain_Verresen_2017} and boundary structures. In gapped systems, these string correlations typically exhibit long-range order, while in gapless systems they may display algebraic decay depending on the interplay between symmetry and critical degrees of freedom~\cite{Jones_2019, Verresen_2021, Thorngren_2021}. However, identifying the appropriate string order parameters generally requires prior knowledge of the underlying symmetry structure, making their construction model-dependent. 

On the other hand, quantum entanglement has been widely used to characterize quantum phases of matter~\cite{Amico_2008, Calabrese_2004, Kitaev_2006, Levin_2006, Li_2008}. However, entanglement is not the only quantum resource that captures quantum complexity. In particular, entanglement measures primarily quantify correlations between subsystems and do not fully reflect the internal structure of quantum states. Non-stabilizerness, or quantum magic, provides a complementary perspective by quantifying the extent to  which a quantum state deviates from the stabilizer manifold and from efficient classical simulability under Clifford dynamics, as characterized by the Gottesman-Knill theorem~\cite{gottesman_1997, Aaronson_2004, Sergey_2005, Veitch_2014, Howard_2017, SRE_Leone_2022}. 

In recent years, several measures of non-stabilizerness have been proposed~\cite{Howard_2017,Veitch_2014, Bravyi_2019, SRE_Leone_2022}. However,  many of them remain notoriously difficult to evaluate for many-body ground states due to the optimization problems inherent in their definitions. A particularly useful proposal is the stabilizer R\'enyi entropy (SRE), which provides a computable measure of quantum magic based on the distribution of Pauli-string expectation values~\cite{SRE_Leone_2022}. The SRE and related magic measures have since been explored in a variety of many-body and hybrid quantum systems~\cite{SRE_Leone_2022, lami2023, Haug_2023, Tarabunga_2023, ding2025, Viscardi_2026, nehra2025, crew_2025}. 

In this work, we studied the stabilizer R\'enyi entropy and the Pauli spectrum in one-dimensional symmetry-protected topological phases, including gapped SPT, non-intrinsically gapless SPT, and intrinsically gapless SPT. Using matrix product state methods and perfect Pauli-string sampling~\cite{lami2023} in large system size, we showed that the stabilizer R\'enyi entropy provides a coarse diagnostic of phase transitions. In all the models studied here, the SRE exhibits an extremum near the transition point, reflecting a redistribution of Pauli-string weights in the ground-state wave function. However, the SRE alone cannot specify the underlying SPT phase. 
On the other hand, by the exact enumeration of Pauli-string weights in small system size, we found that the Pauli spectrum displays a characteristic crossing structure at the transition between different SPT phases. This crossing reflects the exchange of dominant Pauli-string correlations associated with the non-local string order parameters of the two phases. Therefore, although the SRE cannot distinguish the SPT phases, the Pauli spectrum can reveal how the relevant string-order sectors are exchanged across the transition. In this sense, the Pauli spectrum provides a diagnostic of topological phase transitions.

We further showed that the origin of this crossing structure depends on the nature of the SPT phases. For the gapped SPT model and the non-intrinsically gapless SPT model, the crossing can be understood from a local-unitary duality. This duality maps the Pauli-string correlations between the two phases and explains the symmetric structure of the Pauli spectrum. By contrast, in the intrinsically gapless SPT model, such a local unitary mapping is absent. Instead, we found that the Pauli-spectrum mapping can be generated by a non-invertible Kramers-Wannier-type duality transformation. 

The paper is organized as follows. In Sec.\ref{sec:SRE}, we introduce the stabilizer R\'enyi entropy. We discuss the SRE and Pauli spectrum of cluster SPT and $\alpha-$chain model in Sec.\ref{sec:cluster_model}. This section shows that although the SRE cannot distinguish SPT phases, the Pauli spectrum can show a non-local string order parameter crossing to indicate the SPT phases. We also indicate that this crossing originates from the local unitary duality mapping. In Sec.\ref{sec:igSPT}, we compute the SRE and Pauli spectrum of intrinsically gapless SPT model~\cite{Li_2024,Li_2025}. The Pauli spectrum also shows a crossing at the SPT transition point. Although there is no local unitary duality map in this model, we argue that the crossing originates from the non-invertible duality transformation. We conclude our paper in Sec.\ref{sec:conclusion}.

\section{\label{sec:SRE}Stabilizer Rényi Entropy}

In this work, we use the SRE as a measurement of nonstabilizerness for a pure state~\cite{SRE_Leone_2022}. %{\color{red}[PYC: it can also computible for mixed state, maybe we should change to "a generic state"?]}. 
For a quantum system consisting of N qubits, the pure state $|\psi\rangle$ distribution over the $4^N$ Pauli strings $P_N = \{\sigma^0, \sigma^1, \sigma^2, \sigma^3\}$ is defined as
\begin{equation}
\label{eq:M2_pdf}
    p_\psi(\boldsymbol{\sigma})=|\chi(\boldsymbol{\sigma})|^2_\psi = \frac{|\bra{\psi}\boldsymbol{\sigma}\ket{\psi}|^2}{2^N},
\end{equation}
where $\boldsymbol{\sigma} = \prod_{j=1}^{N}\sigma_{j}\in P_N$, and $\{\sigma^\beta\}_{\beta=0}^{3}$ are the Pauli matrices with $\sigma^0 = \mathds{1}$. Note that $\sum_{\boldsymbol{\sigma}\in P_N}{p_\psi(\boldsymbol{\sigma})} = tr|\psi\rangle\langle\psi|^2=1$. Since $p_\psi(\boldsymbol{\sigma})\geq0$ and sum to one, $\{p_\psi(\boldsymbol{\sigma})\}$ is a probability distribution. Here, we refer to this probability distribution as the Pauli spectrum, omitting the normalization factor $2^N$, i.e., Pauli spectrum is defined directly by the squared amplitudes of the Pauli string expectation values $|\bra{\psi}\boldsymbol{\sigma}\ket{\psi}|^2$~\cite{Bera_2025, Turkeshi_2025}.
Furthermore, the $\alpha$-SRE is defined as the ordinary $\alpha$-Rényi of this discrete distribution
\begin{equation}
\label{eq:M2}
    M_\alpha(|\psi\rangle) = (1-\alpha)^{-1}\log{\sum_{\boldsymbol{\sigma}\in P_N}{p_{\psi}(\boldsymbol{\sigma})^{\alpha}}} - N\log{2},
\end{equation}
where $-N\log{2}$ is introduced as shift for convenience. 

The SRE characterizes resources beyond Clifford operations, 
where the underlying Clifford group is generated by the Hadamard, S, and CNOT gates.
The Clifford group is the normalizer of the Pauli group, which means Clifford operations map Pauli operators to Pauli operators, $\text{Cliff}_N:=\{U \in U(2^N)|UP_NU^{\dagger}=P_N\}$. Because of this property, Clifford circuits are free operations, and stabilizer states, $\text{Stab}_N:=\{U\ket{0}^{\otimes N}|U\in \text{Cliff}_N\}$, have zero magic. Note that any stabilizer state is defined by stabilizer group S, subgroup of Pauli group and we have $S(\psi) = \{P\in P_N | P\ket{\psi}=\ket\psi\}$.
%If all terms of the spin Hamiltonian are commute with each other, the ground state is obviously a stabilizer state with zero SRE. [PYC:check this statement].
If a spin Hamiltonian consists entirely of mutually commuting Pauli operators, its constituent terms share a common eigenbasis. The ground state is therefore a simultaneous eigenstate of these operators, which, by definition, constitutes a stabilizer state with vanishing SRE. This makes SRE a meaningful resource-theoretic measure of quantum magic (nonstabilizerness), satisfying several desirable properties:
(i) Faithfulness –  SRE vanishes if and only if the state is a stabilizer state; $M_{\alpha}(|\psi\rangle) = 0$ iff $|\psi\rangle\in$STAB, otherwise $M_{\alpha}(|\psi\rangle) > 0$.
(ii) Invariance under Clifford unitaries – applying allowed Clifford operations does not change the value of SRE; $U\in \text{Cliff}_N$: $M_{\alpha}(U|\psi\rangle) = M_{\alpha}(|\psi\rangle)$.
(iii) Additivity – the SRE of a product state is simply the sum of the SREs of its components; $M_{\alpha}(|\psi\rangle\otimes|\phi\rangle) = M_{\alpha}(|\psi\rangle) + M_{\alpha}(|\phi\rangle).$ 

Stabilizer R\'enyi entropy can also be extended to mixed-state extensions via convex-roof constructions, which preserve monotonicity for $\alpha\geq 2$~\cite{Leone_2024}. Since all states considered in this work are pure ground states, we use the pure-state definition throughout.

Similar to entanglement entropy, this quantity provides a new perspective for studying condensed matter systems~\cite{SRE_liviero_2022, hoshino2025, hoshino2025TD}.
To efficiently compute SRE to large system size, we use perfect Pauli-string sampling~\cite{lami2023} with $\alpha=2$, $M_2$. This algorithm does not suffer from autocorrelation issues in sampling and is easier to do parallel calculation. The DMRG/MPS calculations were performed using ITensor.jl~\cite{itensor, itensor-r0.3}

\section{\label{sec:cluster_model}Cluster Model and gapless SPT}

\subsection{\label{sec:cluster_SPT}Cluster SPT Model}
We first consider the one-dimensional cluster SPT model~\cite{Robert_2001, Nielsen_2006, Son_2011} with Hamiltonian
\begin{equation}
\label{eq:Cluster_SPT}
    H=-(1-h)\sum_iZ_{i-1}X_iZ_{i+1} - h \sum_iX_i,
\end{equation}
where the $X_i$ and $Z_i$ are the Pauli matrices $\{ \sigma^{i},i=1,2,3.\}$. The first term realizes a nontrivial SPT Hamiltonian, while the second term is a symmetry-preserving perturbation. With the $\mathbb{Z}_2^o\times\mathbb{Z}_2^e$ symmetry, we have the symmetry operators, $U^e=\prod_{i\in even}X_i$ and $U^o=\prod_{i\in odd}X_i$. This model hosts a non-trivial SPT phase for $h<\frac{1}{2}$ and a trivial paramagnetic phase for $h>\frac{1}{2}$~\cite{achain_Verresen_2017}.

In the limit $h=0$, the ground state is the cluster state~\cite{zeng2018}, given by 
\begin{equation}
|\psi\rangle = U_{CZ} |+\cdots+\rangle,
\end{equation}
where $U_{CZ} = \prod_{i,i+1}CZ_{i,i+1}$, and the controlled-Z gate acts as $CZ_{i,i+1}|a\rangle_i|b\rangle_{i+1}=(-1)^{ab}|a\rangle_i|b\rangle_{i+1}$, with $\{a,b\}\in{0,1}$.
Equivalently, $CZ_{i,i+1}=\frac{(1+Z_i+Z_{i+1}-Z_iZ_{i+1})}{2}$. Since all terms in the Hamiltonian commute, the ground state is a stabilizer state and therefore has zero SRE.
The cluster Hamiltonian at $h=0$ is related to the trivial paramagnet at $h=1$ by the unitary transformation $U_{CZ}$, which satisfies $U_{CZ}X_iU_{CZ}^{\dagger}=Z_{i-1}X_{i}Z_{i+1}$. This operator is often referred to as the cluster entangler~\cite{Chen_2013, Else_2013, Tantivasadakarn_2023}. Although the global operator $U_{CZ}$ preserves the $\mathbb{Z}_2^o\times\mathbb{Z}_2^e$ symmetry, the individual two-site gates $CZ_{i,i+1}$ do not commute with the local symmetry operators~\cite{Son_2011, Tantivasadakarn_2024}. From the perspective of the finite-depth quantum circuit, the $U_{CZ}$ can connect two distinct (SPT) phases protected by this symmetry, i.e., the local two-site $CZ_{i,i+1}$ explicitly breaks this symmetry. 

\begin{center}
\begin{figure}[h]
    \begin{subfigure}[b]{0.2\textwidth}
        \centering
        \includegraphics[width=\linewidth]{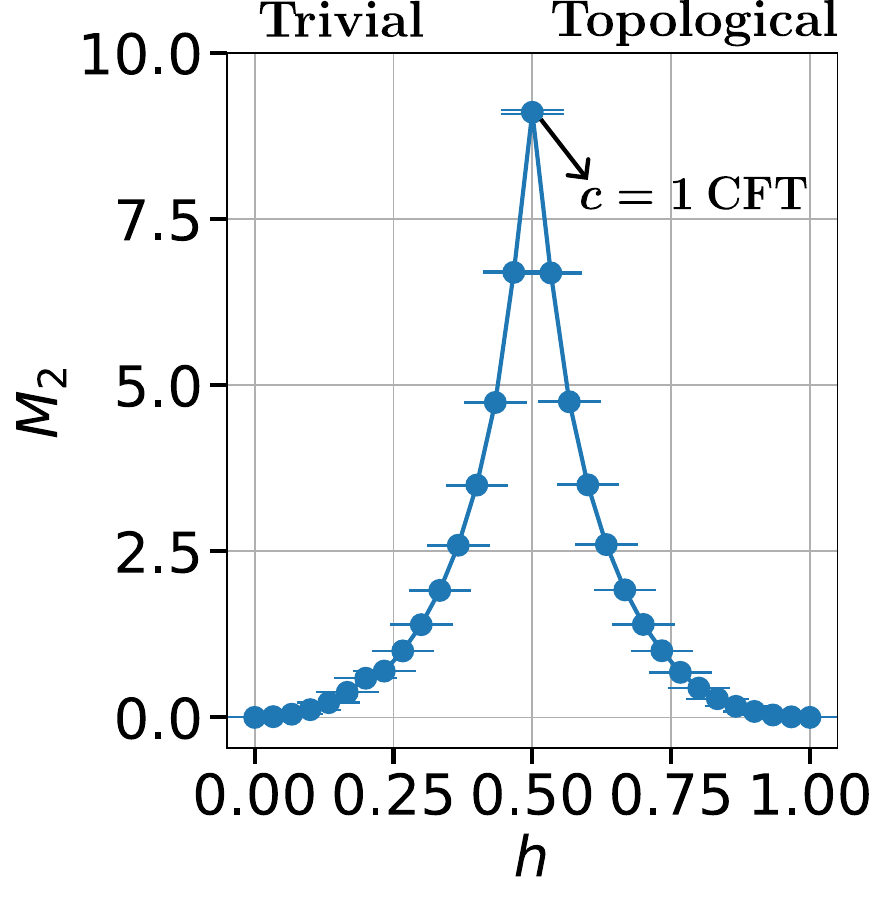}
        \caption{}
    \end{subfigure}
    \begin{subfigure}[b]{0.27\textwidth}
        \centering
        \includegraphics[width=\linewidth]{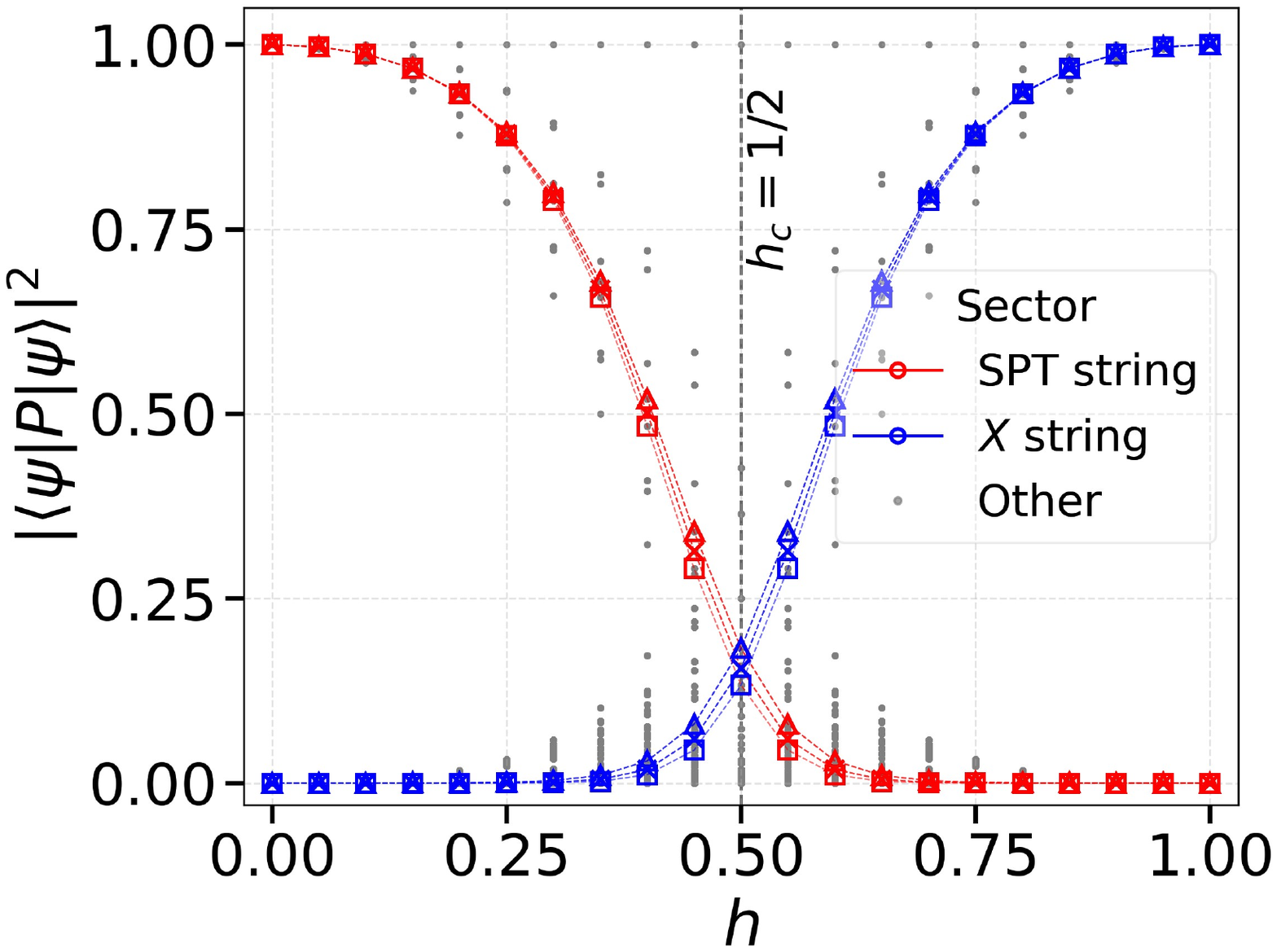}
        \caption{}
    \end{subfigure}
    \caption{(a)$M_2$ of the cluster SPT for $L=30$ under periodic boundary condition(PBC), bond dimension $\chi=50$ and $N_{\text{Samp}}=10^5$. (b)Pauli spectrum for the cluster model with \(L=8\). Red symbols denote SPT non-local string order operators while blue symbols denote trivial non-local string operators. Different marker shapes correspond to different string lengths, defined in Eqs.~\eqref{eq:nls}\eqref{eq:ls}, such as $\triangle: |n-m|=1$, $\times: |n-m|=2$, and $\square: |n-m|=3$. Gray dots show the remaining Pauli-string weights.}
    \label{fig:SRE_PSS_Cluster_SPT}
\end{figure}
\end{center}

$U_{CZ}$ maps the Hamiltonian with the parameter from $h$ to $(1-h)$.
The phase transition occurs at the self-dual point, $h=\frac{1}{2}$. 
Since the $U_{CZ}$ is a Clifford operation that preserves the stabilizer nature (and thus the SRE) of a state, the SRE must be symmetric at the self-dual point $h=1/2$ as shown in Fig.\ref{fig:SRE_PSS_Cluster_SPT}(a).
%However, since $U_{CZ}$ is not an element of the Pauli group, self-duality does not imply the ground state remains a stabilizer state. 
Here we use perfect Pauli sampling algorithm~\cite{lami2023} to compute the SRE by the system size up to $L=30$. The $M_2$ exhibits a peak near the transition point, reflecting a more uniform distribution over Pauli strings. Nevertheless, SRE alone cannot distinguish between the trivial and SPT phase, because $U_{CZ}$ is a Clifford operation satisfying $U_{CZ}P_NU^{\dagger}_{CZ}=P_N$, and thus preserves the Pauli group. Consequently, the ground states in both limiting phases have zero $M_2$.

To resolve this, we consider the full probability distribution of Pauli strings, which we refer to as the Pauli spectrum in Eq.\eqref{eq:M2_pdf}. Due to the duality transformation by $U_{CZ}$, as discussed in Ref.~\cite{Son_2011}, the non-local string order parameter of in the SPT phase,
\begin{align}
\label{eq:nls}
\lim_{|n-m|\rightarrow\infty}\langle Z_{m-1}Y_{m}X_{m+1}\cdots X_{n-1}Y_{n}Z_{n+1}\rangle\sim O(1),
\end{align}
is mapped to the long-range order in the trivial phase, 
\begin{align}
\label{eq:ls}
\lim_{|n-m|\rightarrow\infty}\langle X_mX_{m+1}\cdots X_{n-1}X_{n}\rangle\sim O(1).
\end{align}
As shown in Fig.\ref{fig:SRE_PSS_Cluster_SPT}(b), the Pauli spectrum exhibits a crossing behavior analogous to an order parameter, signaling the phase transition. The crossing originates from the duality transformation, which maps the non-local string order in the SPT phase to trivial correlation. This reflects the switching of dominant non-local string order parameters across the transition, which is captured in the Pauli spectrum.

\subsection{Cluster Ising Model}
\begin{figure*}[t]
    \begin{subfigure}[b]{0.33\textwidth}
        \centering
        \includegraphics[width=\linewidth]{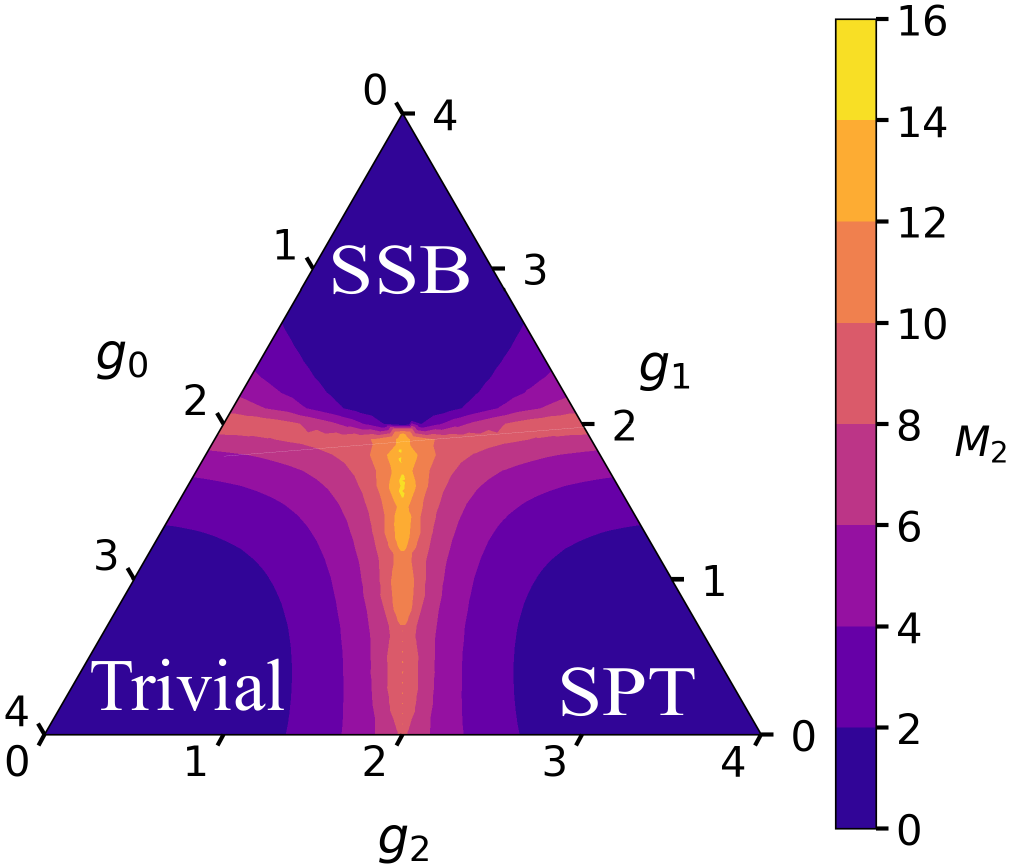}
        \caption{}
    \end{subfigure}
    \begin{subfigure}[b]{0.29\textwidth}
        \centering
        \includegraphics[width=\linewidth]{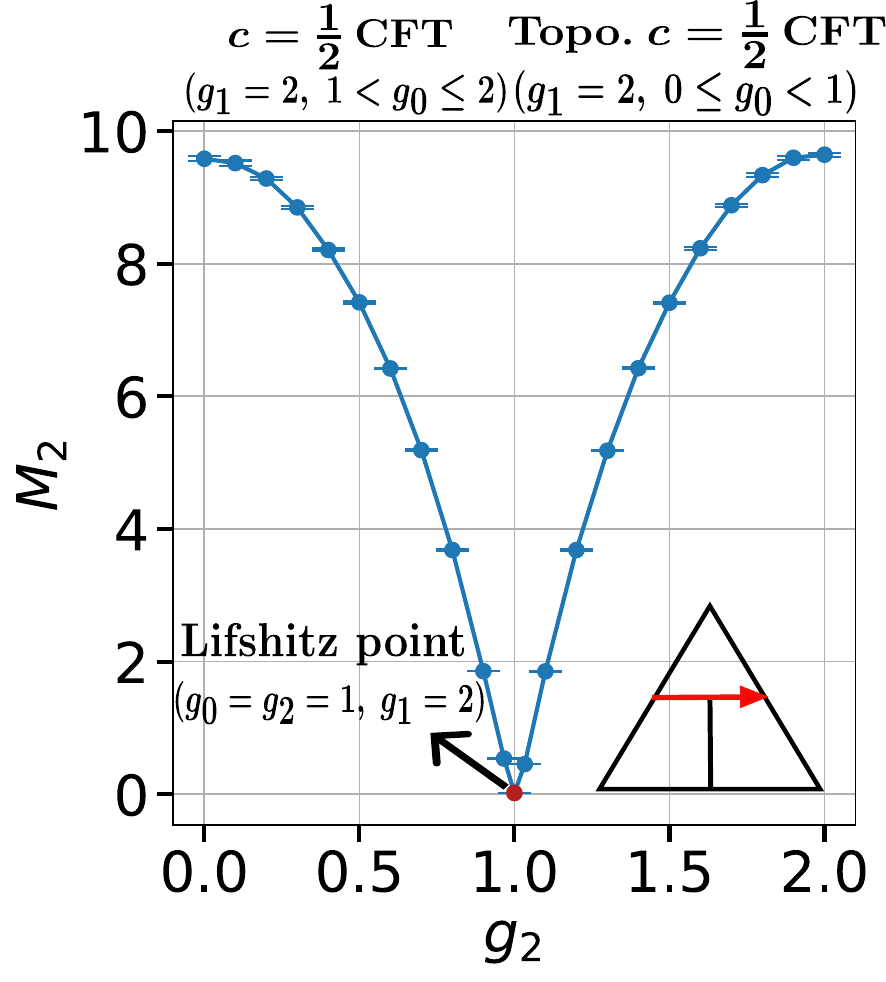}
        \caption{}
        %\label{fig:SRE_phase}
    \end{subfigure}
    \begin{subfigure}[b]{0.36\textwidth}
        \centering
        \includegraphics[width=\linewidth]{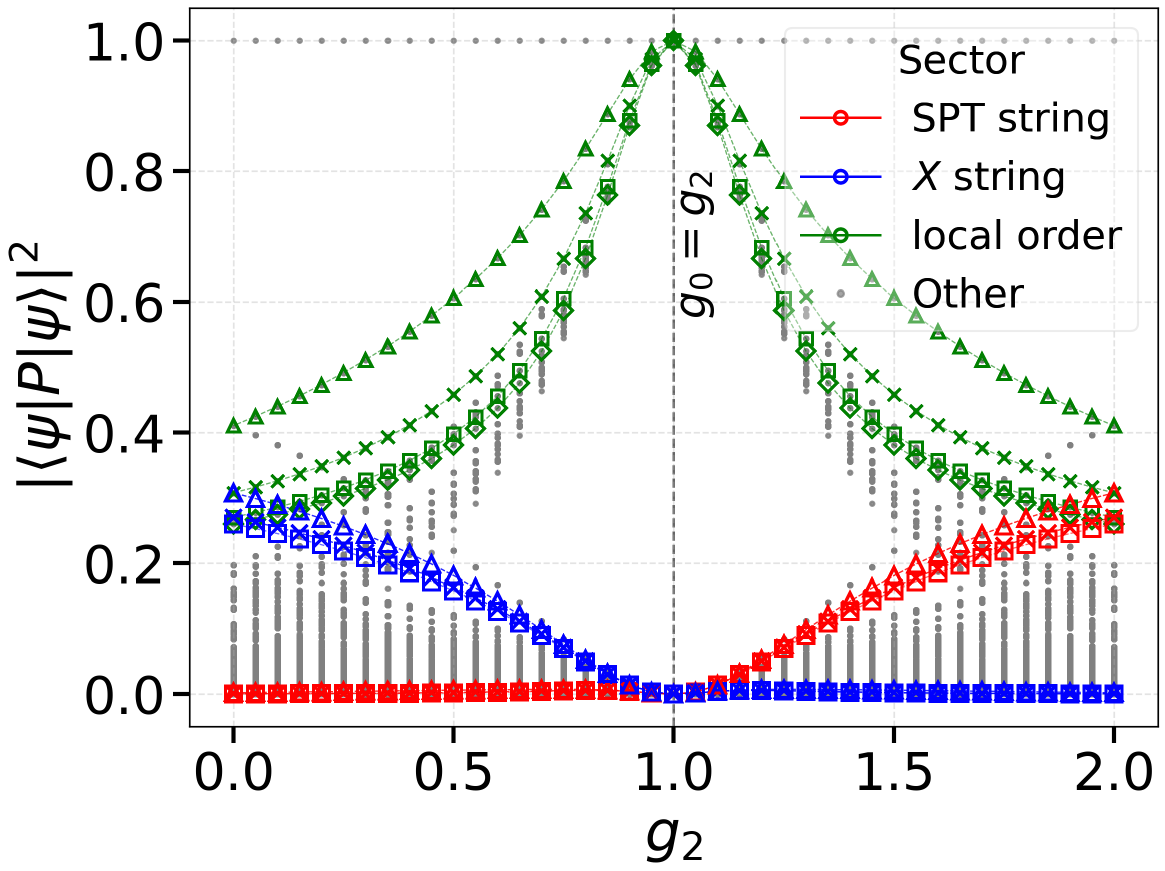}
        \caption{}
    \end{subfigure}
    \caption{
    (a)$M_2$ of the cluster Ising model with PBC, bond dimension $\chi=50$ and $N_{\text{Samp}}=10^5$. (b)$M_2$ along $g_1=2$. (c)Pauli spectrum along $g_1=2$ with $L=8$. Red symbols denote SPT non-local string order operators while blue symbols denote trivial non-local string operators. Green symbols denote SSB local order parameters. Different marker shapes correspond to different string lengths, defined in Eqs.~\eqref{eq:ls}\eqref{eq:nlg}, such as $\triangle: |n-m|=1$, $\times: |n-m|=2$, $\square: |n-m|=3$, and $\diamond: |n-m|=4$. Gray dots show the remaining Pauli-string weights. Note that the X-string is a power law function instead of $O(1)$.}
    \label{fig:SRE_phase_AC}
\end{figure*}

To investigate the symmetry-enriched criticality using SRE and Pauli spectrum, we extend the cluster SPT model in Eq.\eqref{eq:Cluster_SPT} to cluster Ising model by adding a $\mathbb{Z}_2$ spontaneous symmetry breaking(SSB) term in the Hamiltonian. The Hamiltonian is given by
\begin{equation}
\label{eq:AC_model}
    H=\sum_i g_2Z_{i-1}X_iZ_{i+1} - g_1Z_iZ_{i+1} - g_0X_i.
\end{equation}
To present the phase diagram as shown in Fig.~\ref{fig:SRE_phase_AC}(a), we set  $g_0+g_1+g_2=4$.

This model emerges from a stacking $\alpha$ Kitaev chains model via the Jordan-Wigner transformation with a switch of $X_i$ and $Z_i$~\cite{achain_Verresen_2017, achain_Choi_2024, achain_zhong_2025}. These two models are equivalent in open boundary condition. The spin model preserves $\mathbb{Z}_2 \times \mathbb{Z}^T_2$ symmetry with spin-flip symmetry $P=\prod_iX_i$ and time-reversal symmetry $\mathcal{T}=\mathcal{K}$ (complex conjugation). 

Without the SSB perturbation, $g_1=0$, the model goes back to cluster SPT model in Eq.\eqref{eq:Cluster_SPT} with a sign different on $g_2$. At the transition point, $g_0=g_2$ and $g_1=0$, the system shows $c=1$ CFT~\cite{Verresen_2018, achain_Choi_2024}. As shown in Fig.\ref{fig:SRE_phase_AC}(a) and Fig.\ref{fig:samp_check}(c), along the transition point, $g_0=g_2$, with the SSB term $g_1$ increasing, the criticality disappears when the SSB term dominates and a SSB ground state degeneracy emerges starting from $(g_0,g_1,g_2)=(1,2,1)$. This point is a $z=2$ Lifshitz transition point.
Actually, this Lifshitz point is a tricritical point. The appearance of this Lifshitz point implies a symmetry enrich criticality transition. As shown in Fig.\ref{fig:SRE_phase_AC}(b), along $g_1=2$, there are two topological distinct Ising CFT phases.
%there are two symmetry enrich Ising CFT phases.\cite{Verresen_2021}. 
At $g_2=0$, it is a trivial Ising criticality, while at $g_2=2$, generalized cluster Ising ~\cite{Verresen_2018}.

The SRE calculation with L=30 is shown in Fig.\ref{fig:SRE_phase_AC}(a)(b) and Fig.\ref{fig:samp_check}(c) with uncertainty in Fig.\ref{fig:samp_check}(b). We find that the SRE identifies the phase boundaries of this model. Moreover, the Hamiltonian has a self-duality generated by $P_ZU_{CZ}$ and $P_Z=\prod_iZ_i.$ This unitary exchanges the transverse-field term and the cluster term with a minus sign, 
\begin{align*}
    &(P_ZU_{CZ})X_i(P_ZU_{CZ})^\dagger=-Z_{i-1}X_iZ_{i+1},\\
    &(P_ZU_{CZ})Z_{i-1}X_iZ_{i+1}(P_ZU_{CZ})^\dagger=-X_i,
\end{align*}
while leaving the Ising term invariant. Therefore $(P_ZU_{CZ})H(g_0,g_1,g_2)(P_ZU_{CZ})^\dagger=H(g_2,g_1,g_0)$ and the self-dual line is $g_0=g_2$. Correspondingly, $M_2$ shows a symmetric structure about this line. However, as shown in Fig.\ref{fig:samp_check}(c), the SRE can only signals the transition, and does not distinguish the two distinct topological critical phases. At the tricritical point, $(g_0,g_1,g_2)=(1,2,1)$, the ground state is the Greenberger-Horne-Zeilinger (GHZ) state~\cite{achain_Choi_2024}, which is a stabilizer state and therefore has vanishing SRE. Similarly, in the three limiting regions dominated by $g_0, g_1$, and $g_2$, all terms in the Hamiltonian commute, so the ground states are stabilizer states with zero $M_2$. Thus, in this model, the SRE serves as a useful diagnostic of phase boundaries but not for the phases.

%{\color{red}Since the Hamiltonian also dual by $PU_{CZ}$ operator with self-dual point, $g_0=g_2$, the $M_2$ shows a symmetric result. In the three limit of Hamiltonian, since all terms of Hamiltonian are commute, so the ground states are stabilizer state and show zero $M_2$ in Fig.\ref{fig:SRE_phase_AC}(a).
%[PYC: First need to express the transformation by $PU_{CZ}$ operator and show that there is a self-dual point at $g_0=g_2$. Second, express in details about the three limits of the Hamiltonian.]
%} In the region of $g_2=0$, the model reduces to the transverse field Ising model, which is related by the standard Kramers-Wannier duality. On a periodic chain, this duality is not an ordinary unitary conjugation but a non-invertible Pauli-to-Pauli intertwiner. We will use this viewpoint more explicitly in Sec.\ref{sec:igSPT} for the intrinsically gapless SPT model.
%Furthermore, the SRE can specify the phase boundary of this model, as shown in Fig.\ref{fig:SRE_phase_AC}(a-c). Since ground state at the tricritical point, $(g_0,g_1,g_2)=(1,2,1)$, is the Greenberger-Horne-Zeilinger (GHZ) state~\cite{achain_Choi_2024}, so it has vanishing SRE.  

By contrast, the Pauli spectrum resolves the switching of the non-local string order along the symmetry-enriched Ising critical line, as shown in Fig.\ref{fig:SRE_phase_AC}(c).
Here the symmetry associated with the non-local order parameter comes from the $Z_2$-enriched Ising CFT which leads to the power law behavior of the  non-local order parameter~\cite{Verresen_2021}:
\begin{align}
\label{eq:nlg}
\lim_{|n-m|\rightarrow\infty}\langle Z_{m-1}Y_{m}X_{m+1}\cdots X_{n-1}Y_{n}Z_{n+1}\rangle\sim \frac{1}{|n-m|^{2\Delta_g}}.
\end{align}

%Note that the scaling dimensions of these operators serve as an order parameter encoding topological invariant and central charge~\cite{Jones_2019}
Note that the scaling dimensions of these non-local string operators serve as a diagnostic at criticality, encoding both the bulk topological invariant and the central charge of the CFT, i.e., $\Delta_g (\omega,c)$ with $\omega$ being the topological invariant and $c$ being the central charge~\cite{Jones_2019}.
In Fig.\ref{fig:SRE_phase_AC}(c),
the probability weight of the non-local string order parameter with difference length $|n-m|$ shows a descent behavior as a function of  $|n-m|$ with the parameter $g_1=2$. Due to the $P_ZU_{CZ}$ duality, the Pauli spectrum exhibits a crossing structure corresponding to the switching of string order sectors, similar as cluster SPT in Sec.\ref{sec:cluster_SPT}. 
Since the critical line along $g_2$ with fixed $g_1=2$ is the phase boundary of SSB, the associated local order parameter, $\langle Z_{i}Z_{j}\rangle$ also shows in Fig.\ref{fig:SRE_phase_AC}(c). 

Together, the two cluster-type models  in Eq.~\eqref{eq:Cluster_SPT} and Eq.\eqref{eq:AC_model}, show that the SRE can show extremal value at phase boundary and Pauli spectrum  capture the switching of dominant non-local string order parameters across transition due to the $U_{CZ}$ and $P_ZU_{CZ}$ dualities.

%{\color{red}[PYC: should we say the first derivative of the SRE is discontinuous at the transition point?]}{\color{blue}[YLL: I am not sure whether emphasizing a discontinuity in the first derivative of the SRE would provide additional useful information. It may be safer to state only that the SRE shows an extremum near the transition. The Pauli spectrum already gives a more direct explanation of the crossing and switching of the dominant string-order sectors.]}

\section{\label{sec:igSPT}Intrinsically gapless SPT}

In the previous examples, the crossing structure in the Pauli spectrum can be understood from the SPT entangler, namely a locality-preserving unitary that relates different SPT phases by changing the symmetry charge carried by nonlocal string operators~\cite{Saranesh_2025}. 
We now examine an intrinsically gapless SPT (igSPT) model.
%By definition, such a model possesses an inherent low-energy anomaly and fundamentally cannot be built or trivialized by simply stacking or decoupling gapped SPT components.
In contrast to non-intrinsically gapless SPT, an igSPT phase possesses an inherent low-energy anomaly and fundamentally cannot be trivialized by stacking with a gapped SPT phase.
Consequently, the conventional local-unitary picture is not applicable. Nevertheless, we find that the Pauli spectrum still preserves a distinct duality structure. 
We show that this structure comes from a Kramers-Wannier(KW)-type non-invertible map rather than from a symmetric local unitary transformation.

Following Refs.~\cite{Thorngren_2021, Li_2024, Li_2025}, we first review the Kennedy-Tasaki (KT) construction of the $Z_4^{\Gamma}$. From the construction, one starts from a decoupled Hamiltonian consisting of an XX chain in the $\tau$ sector and an Ising symmetry-breaking chain in the $\sigma$ sector,
\begin{equation}
\label{eq:IsingSSB}
    H_{XX+SSB} = -\sum_{i=1}^{L_{\text{unit}}}{\tau_{i-\frac{1}{2}}^{z}\tau_{i+\frac{1}{2}}^z + \tau_{i-\frac{1}{2}}^{y}\tau_{i+\frac{1}{2}}^y + \sigma_{i-1}^{z}\sigma_i^z}.
\end{equation}

The $\tau$ sector is gapless, while the $\sigma$ sector is in a symmetry-breaking phase. Under the KT transformation, the local terms are mapped as
$$\tau^z_{i-\frac{1}{2}}\tau^z_{i+\frac{1}{2}} \rightarrow \tau^z_{i-\frac{1}{2}}\sigma_i^x\tau^z_{i+\frac{1}{2}},$$
$$\tau^y_{i-\frac{1}{2}}\tau^y_{i+\frac{1}{2}} \rightarrow \tau^y_{i-\frac{1}{2}}\sigma_i^x\tau^y_{i+\frac{1}{2}},$$
and
$$\sigma^z_{i-1}\sigma^z_{i} \rightarrow \sigma^z_{i-1}\tau_{i-\frac{1}{2}}^x\sigma^z_{i}.$$

\begin{figure}[t]
    \begin{subfigure}[b]{0.2\textwidth}
        \centering
        \includegraphics[width=\linewidth]{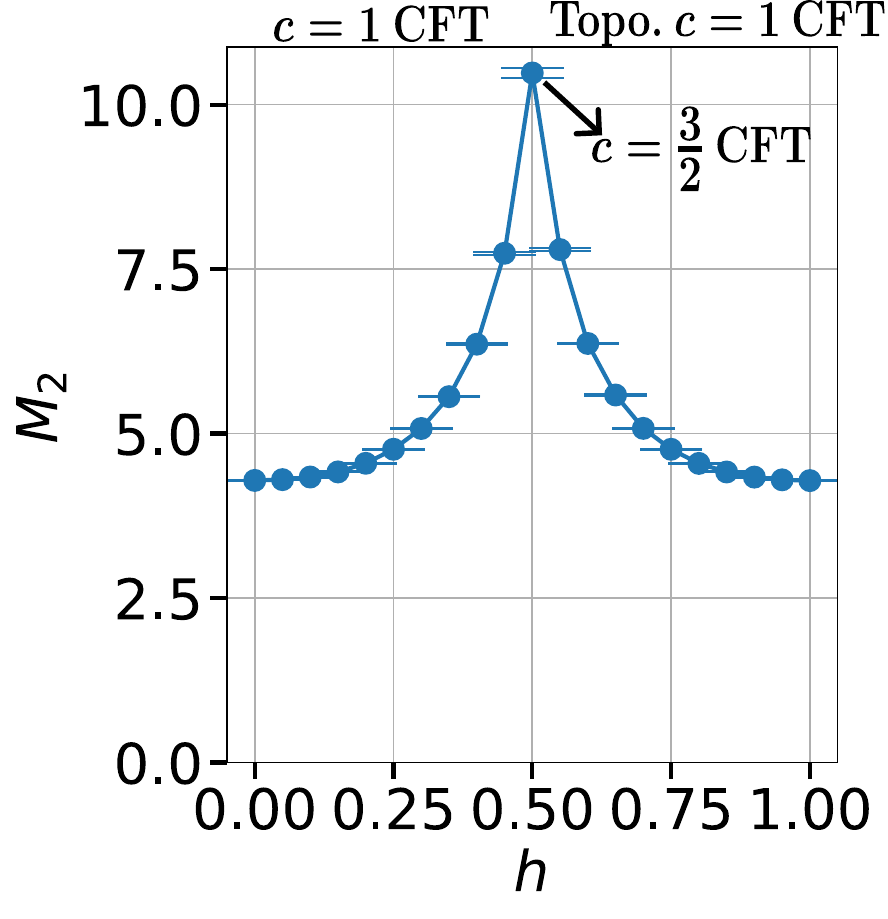}
        \caption{}
    \end{subfigure}
    \begin{subfigure}[b]{0.27\textwidth}
        \centering
        \includegraphics[width=\linewidth]{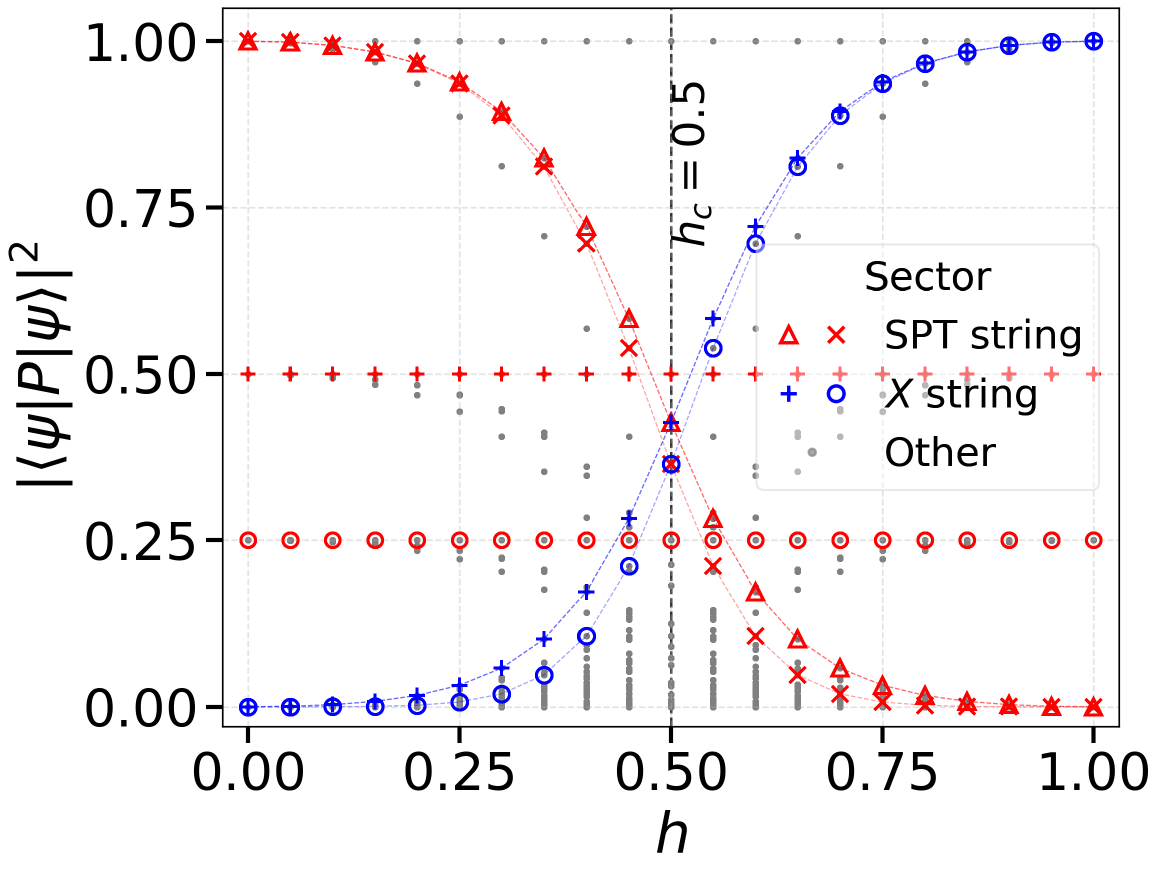}
        \caption{}
    \end{subfigure}
    \caption{(a)$M_2$ of igSPT for $L_{\text{unit}}=16$ (32 spins) with PBC, bond dimension$\chi=100$, $N_{samp}=10^5$. (b)Pauli string weight for $L_{\text{unit}}=4$ (8 spins). Red symbols denote SPT non-local string order operators, defined in Eq.~\eqref{eq:string_op_igSPT}, with $\triangle: |i-j|=1$, and $\times: |i-j|=2$ while blue symbols denote trivial non-local string operators, defined in Eq.~\eqref{eq:trivial_string_op_igSPT}, with $+: |i-j|=1$, and $o: |i-j|=2$. Note that the Red symbols with $+: |i-j|=1$, and $o: |i-j|=2$ is defined in Eq.~\eqref{eq:_op_order}}
    \label{fig:SRE_PSS_igSPT}
\end{figure}

This gives the $\mathbb{Z}_4^\Gamma$ igSPT Hamiltonian studied in Ref.\cite{Li_2024}. 
%In the present work, we take this KT-generated model as our starting point and introduce a symmetry-preserving interpolation between the KT image of the Ising coupling and the transverse-field term in the $\sigma$ chain:
In the present work, we take this KT-generated model as our starting point and introduce a symmetry-preserving interpolation between the transverse-field term ($\sigma^x_i$) and the KT transformation of the original Ising coupling:

\begin{align}
\label{eq:igSPT}
H_{\mathrm{igSPT+pert}}
&= -\sum_{i=1}^{L_{\text{unit}}} \Big(
    \tau_{i-\frac{1}{2}}^{z}\sigma_i^x\tau_{i+\frac{1}{2}}^z
    + \tau_{i-\frac{1}{2}}^{y}\sigma_i^x\tau_{i+\frac{1}{2}}^y \nonumber \\
&\qquad
    + (1-h)\sigma_{i-1}^{z}\tau_{i-\frac{1}{2}}^x\sigma_i^z
    + h\sigma_i^x
\Big).
\end{align}
The Hamiltonian is invariant under the $\mathbb{Z}_4^{\Gamma}$ symmetry generated by $U_\Gamma = U_{\sigma}V_{\tau}$, 
where 
\begin{equation*}
    U_{\sigma}=\prod_{i=1}^{L_{\text{unit}}}\sigma_i^x,
    V_{\tau}=\prod_{i=1}^{L_{\text{unit}}}e^{\frac{i\pi}{4}(1-\tau_{i-\frac{1}{2}}^x)}.
\end{equation*}
In fact, the Hamiltonian has a larger $\mathbb{Z}_2^{\sigma}\times\mathbb{Z}_4^{\tau}$ symmetry. In this work, however, we focus on the $\mathbb{Z}_4^{\Gamma}$ structure relevant to the intrinsically gapless SPT.
The numerical results in Fig.\ref{fig:SRE_PSS_igSPT} show that both the SRE and the Pauli spectrum is symmetric under $h \leftrightarrow (1-h)$.
%a clear $h \leftrightarrow 1-h$ structure. 
This suggests that the two $h$-dependent terms are exchanged by a duality, while the remaining terms are preserved.
In addition, we also find out that there is no invertible operator, which can implement this Pauli duality on the full periodic Hilbert space in this igSPT model in Appendix~\ref{sec:no_inver_dual}. Therefore, the crossing structure found here should not be interpreted as coming from an ordinary local-unitary SPT entangler, but from a non-invertible KW-type duality.

Before constructing this duality, we briefly clarify the role of the KW and KT transformations. On a periodic chain, the Kramers-Wannier transformation is not an ordinary unitary conjugation on the full Hilbert space~\cite{Seiberg_2024_1, Seiberg_2024_2, Parayil_Mana_2024}. It can be written schematically as a unitary Clifford operator and a even parity projection operator
\begin{equation}
\label{eq:KW_trans}
\mathbf{D}=\mathbf{\tilde{D}}\mathbf{\eta},
\end{equation}
where $\mathbf{\tilde{D}}=(\prod_{i=1}^{L-1}e^{i\frac{\pi}{4}X_i}e^{i\frac{\pi}{4}Z_iZ_{i+1}})e^{i\frac{\pi}{4}X_L}$ and $\mathbf{\eta}=\frac{1+P}{2}$. Thus $\mathbf{D}$ is non-invertible. Nevertheless, it acts as a Pauli-to-Pauli map in the intertwining sense. For example, in the Ising chain, $\mathbf{D}X_j = Z_{j-1}Z_j\mathbf{D}$, $\mathbf{D}Z_{j-1}Z_j = X_{j-1}\mathbf{D}$. 
The KT transformation used here has a similar non-invertible character on a ring. It can be represented as
\begin{align}
U_{CZ}\mathbf{D^d}U_{CZ},
\end{align}
where $\mathbf{D^d}=T\mathbf{D^o}\mathbf{D^e}$ is a double KW transformation, with $\mathbf{D^o}$ and $\mathbf{D^e}$ acting on the odd and even sublattices, respectively, or in our case, on the $\tau$ and $\sigma$ spins. $T$ is a lattice translation operator. Because the KW steps contain projectors, the full KT transformation on a periodic chain is non-invertible and should not be regarded as a Clifford unitary circuit. Li-Oshikawa-Zheng\cite{Li_2025} also emphasize that the KT transformation is non-unitary on a ring and is better understood as a non-invertible duality transformation; it becomes unitary only under suitable open-boundary conditions.

We now construct the $h \leftrightarrow (1-h)$ duality using the KT transformation. First we transform the Hamiltonian back to the decoupled model by the KT transformation, then perform the KW transformation only on the $\sigma$ chain, and finally apply KT transformation again. Thus the duality is
$$\mathcal{N} = KT \circ KW^{\sigma} \circ KT.$$
Here $KW^{\sigma}$ denotes the KW transformation acting only on the $\sigma$ sublattice. In the decoupled representation, it exchanges the Ising coupling and the transverse-field term. Therefore,
$$\mathcal{N}H_{\mathrm{igSPT+pert}}(h) = H_{\mathrm{igSPT+pert}}(1-h)\mathcal{N}.$$
This implies that the $h \leftrightarrow (1-h)$ structure is generated by a KW-type non-invertible duality transformation.

We emphasize that $\mathcal{N}$ is not a finite-depth local unitary circuit. It is a non-invertible transformation acting on the relevant Pauli group. 
In Appendix~\ref{sec:KW_Cliffordmap}, we demonstrate the Pauli spectrum mapping by the KW transformation if the system shows a KW duality map with even parity eigenstates. We have also checked the ground state of the Hamiltonian with $0\leq h\leq 1$ and $L_{\text{unit}}=4,8,12,16$ are even parity. As a result, by the Pauli spectrum mapping of KW transformation with even parity state, this non-invertible transformation $\mathcal{N}$ also provides a Pauli spectrum map with the ground state of igSPT model. 

Let us discuss the phase diagram and the non-local string order parameters as first studied in Ref.~\cite{Li_2025}. The phase structure follows naturally from the decoupled representation. Before applying KT, the $\tau$ sector remains an XX chain, while the $\sigma$ sector is a transverse-field Ising chain with Ising coupling $(1-h)$ and transverse field $h$. Therefore, the self-dual point is at $h=\frac{1}{2}$.

For $h<1/2$, the $\sigma$ sector is in the symmetry-breaking phase before KT, and the KT-transformed system realizes the nontrivial igSPT phase. For $h>1/2$, the $\sigma$ sector is in the trivial paramagnetic phase before KT, and the KT-transformed system realizes a trivial gapless phase. At $h=1/2$, the $\sigma$ sector becomes Ising critical, while the $\tau$ sector remains gapless. Thus, the transition has total central charge $c=\frac{3}{2}$ and the other parameter space is $c=1$ free boson CFT coming from the XX chain. 

On the other hand, the non-local string order parameters can also be computed by KT transition from the conventional correlation functions by the decoupled Hamiltonian. The two-point correlation functions and the non-local string order parameters exhibit the scaling:
\begin{equation}
    \langle \sigma^z_{i} \sigma^z_{j} \rangle \rightarrow \left\langle \sigma^z_i \left( \prod_{k = i}^{j - 1} \tau^x_{k + \frac{1}{2}} \right) \sigma^z_j \right\rangle \sim 
    \mathcal{O}(1),
\label{eq:string_op_igSPT}
\end{equation}

\begin{equation}
    \langle \tau^z_{i - \frac{1}{2}} \tau^z_{j - \frac{1}{2}} \rangle \rightarrow\left\langle \tau^z_{i - \frac{1}{2}} \left( \prod_{k = i}^{j - 1} \sigma^x_k \right) \tau^z_{j - \frac{1}{2}} \right\rangle \sim 
    \frac{1}{|i - j|^{2\Delta}}.
\label{eq:_op_order}
\end{equation}
where \(\Delta\) denotes the scaling dimension of the \(\tau^z\) operator in the XX chain. 
Here the symmetry associated with the non-local string order parameter in Eq.~\eqref{eq:string_op_igSPT} comes from the gapped sector, it shows the long range order $O(1)$.
These string order parameters provide the standard diagnostic of the igSPT structure, while the Pauli spectrum gives a complementary operator-space view.

As shown in Fig.\ref{fig:SRE_PSS_igSPT}(a), the $M_2$ is symmetric under $h\leftrightarrow (1-h)$. Since this is a symmetry enriched criticality transition with a higher central charge changing at the transition point, all the $M_2$ are nonzero at the parameter space and reaches a sharp maximum at the self-dual point $h=1/2$ to indicate the phase transition. This enhancement indicates that the Pauli spectrum becomes most broadly distributed near the transition, where the two $h$-dependent Pauli terms are exchanged.

In Fig.\ref{fig:SRE_PSS_igSPT}(b), the non-local string order parameters in Eq.~\eqref{eq:string_op_igSPT} also show O(1) at small perturbation and decrease as perturbation increasing since these order parameters are originated from gapped sector. Furthermore, the same duality of non-invertible operation $\mathcal{N}$ is visible more directly in the Pauli spectrum, with the relative non-local string order parameters for the trivial phase:
\begin{equation}
    \left\langle \prod_{k = i}^{j-1} \sigma^x_k  \right\rangle \sim 
    \mathcal{O}(1).
\label{eq:trivial_string_op_igSPT}
\end{equation}

\section{\label{sec:conclusion}Conclusion}
In this work, we studied the stabilizer R\'enyi entropy and the Pauli spectrum in gapped and gapless symmetry-protected topological phases. We found that the stabilizer R\'enyi entropy can signal phase transitions through extremal behavior, but it does not distinguish different SPT phases. By contrast, the Pauli spectrum provides more detailed information. In all the examples studied here, the Pauli spectrum shows a crossing structure at the transition point, reflecting the exchange of dominant non-local string order parameters between the two phases. For the gapped SPT and non-intrinsically gapless SPT cases, this crossing can be understood from an invertible local-unitary duality. For the intrinsically gapless SPT case, however, the same type of Pauli-spectrum mapping is generated by a non-invertible duality transformation instead of a local unitary.

%These results suggest that the Pauli spectrum gives an persepctive on SPT phases that is not captured by the stabilizer R\'enyi entropy alone.
%{\color{red} These results suggest that the Pauli spectrum reveals a detailed structure of SPT phases that reflects both their non-local string order parameters and their duality mapping to spontaneous symmetry breaking phases—features that remain entirely uncaptured by the stabilizer Rényi entropy alone. According, quantum magic provides a complementary way to characterize gapped and gapless SPT phases.}
These results suggest that the Pauli spectrum gives a persepctive on SPT phases that is not captured by the stabilizer R\'enyi entropy alone. While the stabilizer R\'enyi entropy provides a useful coarse diagnostic of phase transitions, the full Pauli spectrum retains information about which non-local correlations dominate the many-body wave function. This makes the Pauli spectrum a useful probe of the interplay between quantum magic, string order, and duality in both gapped and gapless SPT phases.

\begin{acknowledgments}
The work is supported by National Science and Technology Council of Taiwan under Grants No. NSTC 113-2112-M-007-019, 114-2918-I-007-015. 
We thank Shinsei Ryu for stimulating discussion and acknowledges support from the National Center for Theoretical Sciences, Physics Division.
Ying-Lin Li thanks Bo-Han Lin and Chia-Hsin Chen for valuable discussion.
\end{acknowledgments}

\appendix
\section{Convergence with Sampling Number}

\begin{figure*}[t]
    \begin{subfigure}[b]{0.32\textwidth}
        \centering
        \includegraphics[width=\linewidth]{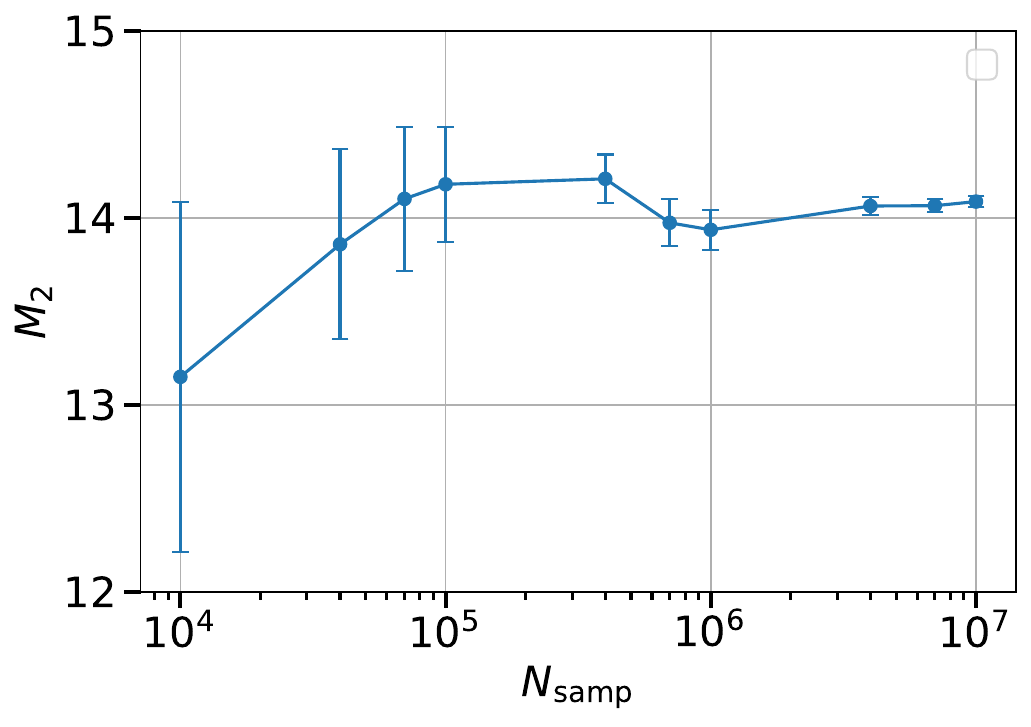}
        \caption{}
    \end{subfigure}
    \begin{subfigure}[b]{0.32\textwidth}
        \centering
        \includegraphics[width=\linewidth]{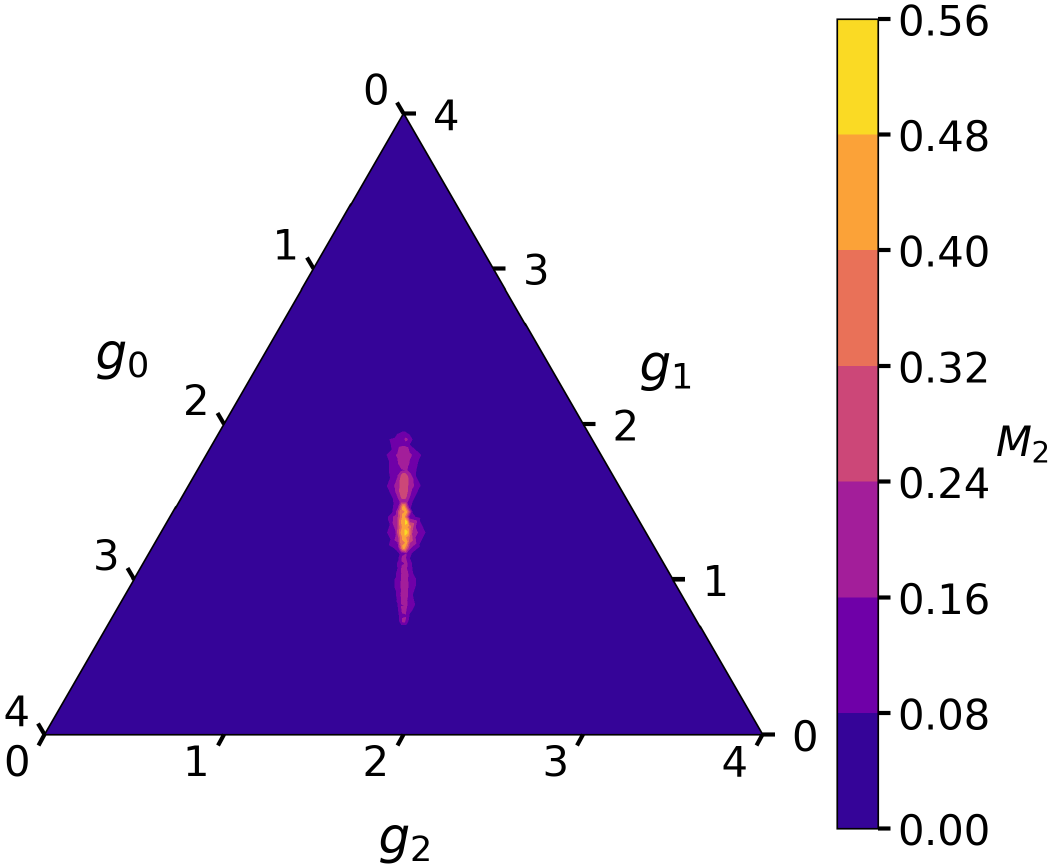}
        \caption{}
    \end{subfigure}
    \begin{subfigure}[b]{0.26\textwidth}
        \centering
        \includegraphics[width=\linewidth]{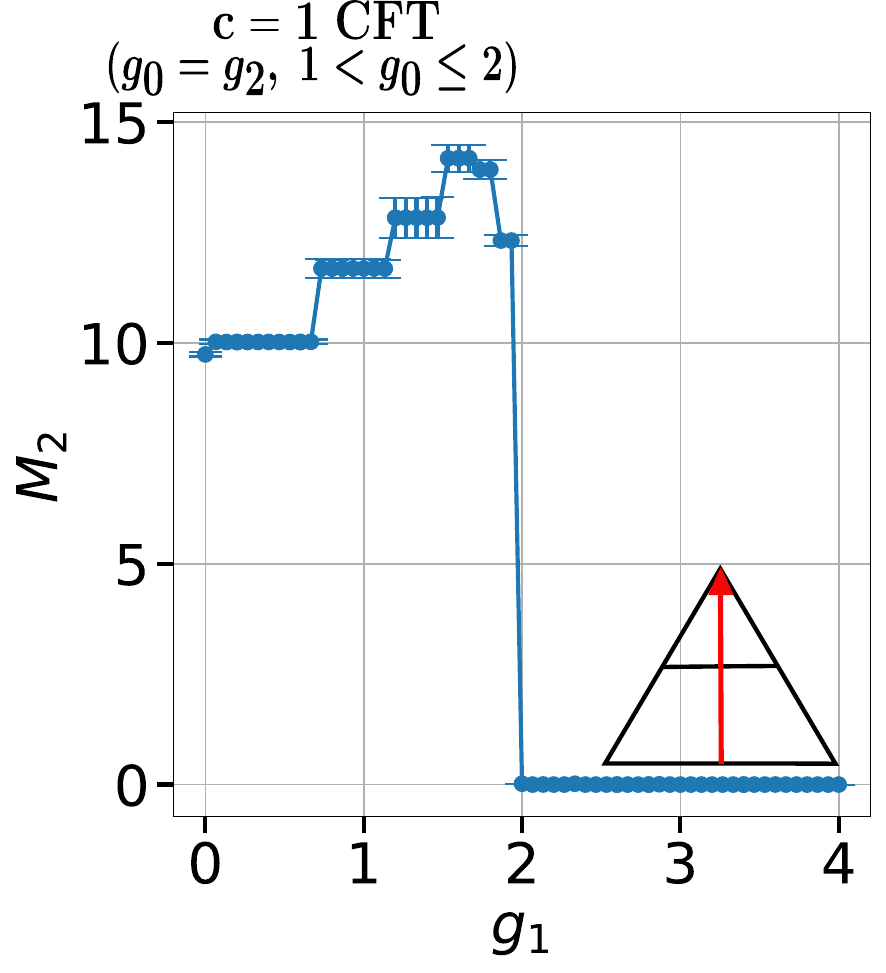}
        \caption{}
    \end{subfigure}
    
    \begin{subfigure}[b]{0.32\textwidth}
        \centering
        \includegraphics[width=\linewidth]{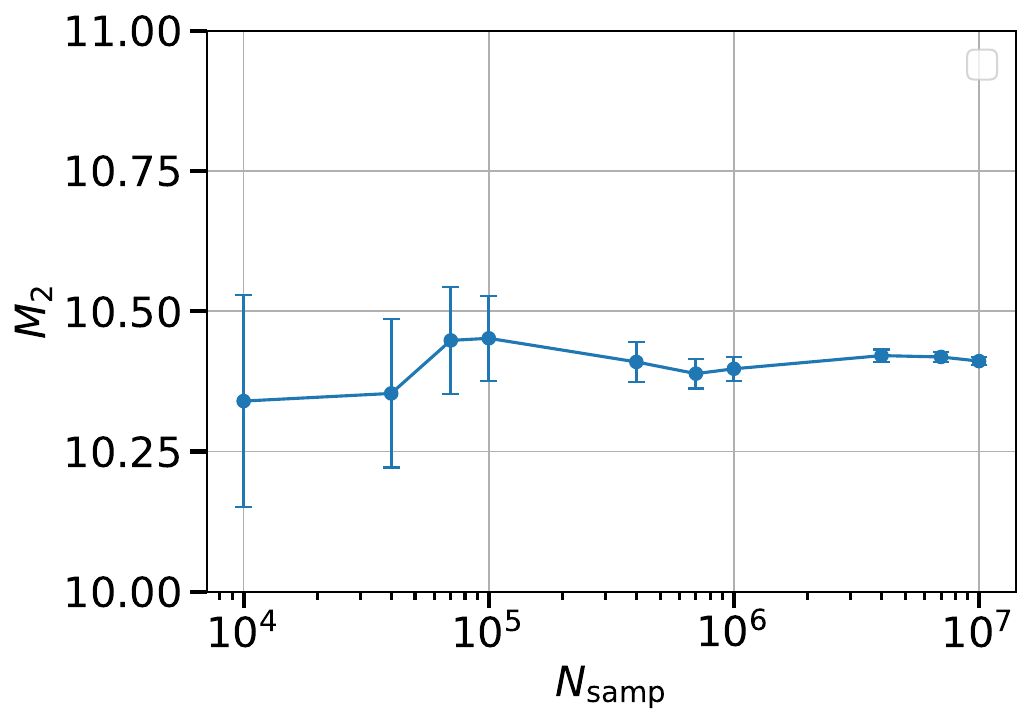}
        \caption{}
    \end{subfigure}
    \begin{subfigure}[b]{0.32\textwidth}
        \centering
        \includegraphics[width=\linewidth]{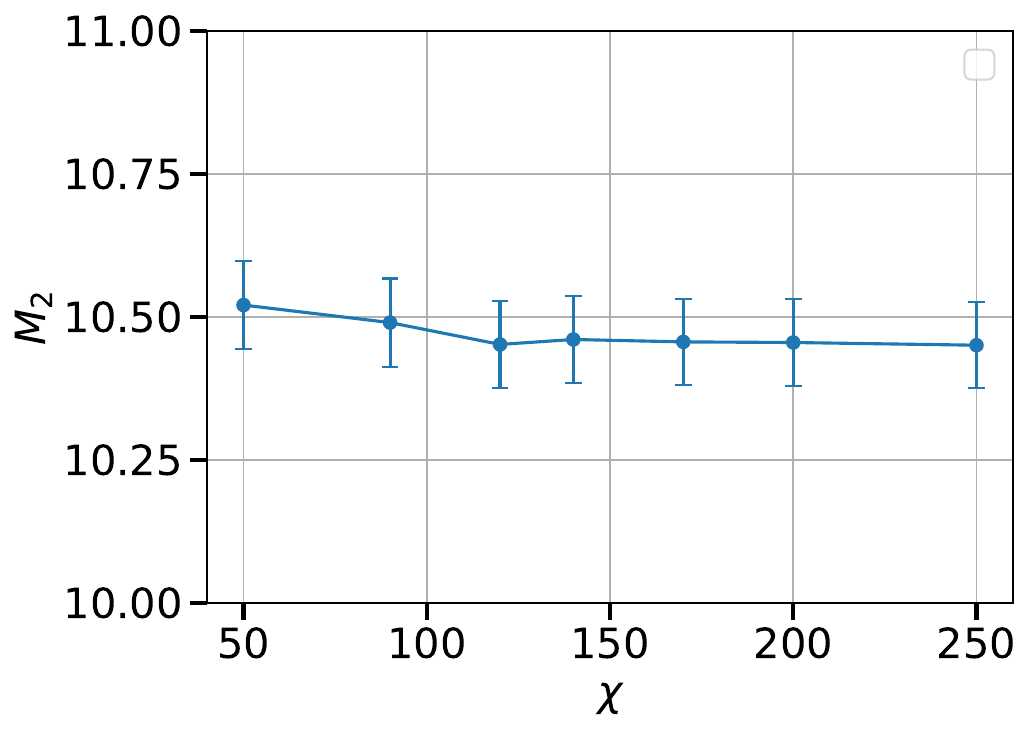}
        \caption{}
    \end{subfigure}

    \caption{(a)Convergence of $M_2$ with respect to number of sampling $N_{\text{Samp}}$ of Cluster-type models with L=30, $\chi=50$ at $(g_0, g_1, g_2)=(1.2,1.6,1.2)$. (b)The error of the SRE sampling calculation of Cluster Ising model with $L=30$, $\chi=50$, and $N_{\text{Samp}}=10^5$. (c)$M_2$ along $g_0=g_2$ of the cluster Ising model with PBC, $\chi=50$ and $N_{\text{Samp}}=10^5$. (d)Convergence of $M_2$ with respect to number of sampling $N_{\text{Samp}}$ of igSPT models with $L=32$, $\chi=100$ at $h=\frac{1}{2}$. (e)The bond dimension check of $M_2$ with $N_{\text{Samp}}=10^5$ of igSPT models with $L=32$ at $h=\frac{1}{2}$.}
    \label{fig:samp_check}
\end{figure*}

Here we discuss the convergence of the MPS perfect-sampling calculations. For the cluster model in eq.\ref{eq:AC_model}, we fixed the bond dimension to $\chi=50$ at $(g_0, g_1, g_2)=(1.2,1.6,1.2)$ and examine the convergence of $M_2$ with respect to the number of samples, as shown in Fig.\ref{fig:samp_check}(a). We choose number of samples $N_{\text{samp}}=10^5$ in this case. We have also checked the bond dimension $\chi$ is large enough. The $M_2$ and the error do not change as $\chi$ increases. The error of the phase diagram of Cluster Ising model is shown in Fig.\ref{fig:samp_check}(b). In Fig.\ref{fig:samp_check}(c), we show $M_2$ along the self-dual line $g_0=g=2$.

For the intrinsically gSPT model in eq.\ref{eq:igSPT}, we use bond dimension $\chi=100$ at $h=\frac{1}{2}$, and the corresponding sampling number convergence is shown in Fig.\ref{fig:samp_check}(d). We choose the sampling number $N_{\text{samp}}=10^5$. We also check the effect of bond dimension in Fig.\ref{fig:samp_check}(e).

\section{\label{sec:no_inver_dual}Obstruction to an invertible Pauli-string duality map on a periodic chain}

We now show that the above duality map of Hamiltonian in eq.\ref{eq:igSPT} cannot be implemented by an invertible operator on the full periodic Hilbert space. Suppose that there exists an invertible operator $U$ realizing the Pauli-string map
$$U\sigma^z_{i-1}\tau^x_{i-\frac{1}{2}}\sigma^z_{i}U^{-1} = \sigma^x_i,$$ while preserving the set of remaining two terms, $$\{\tau^z_{i-\frac{1}{2}}\sigma^x_i\tau^z_{i+\frac{1}{2}}, \tau^y_{i-\frac{1}{2}}\sigma^x_i\tau^y_{i+\frac{1}{2}}\} \leftrightarrow \{\tau^z_{i-\frac{1}{2}}\sigma^x_i\tau^z_{i+\frac{1}{2}}, \tau^y_{i-\frac{1}{2}}\sigma^x_i\tau^y_{i+\frac{1}{2}}\},$$ possible up to lattice translations.

Taking the product over $i$, the first relation gives
$$U\left(\prod_{i}\sigma^z_{i-1}\tau^x_{i-\frac{1}{2}}\sigma^z_{i}\right)U^{-1} = \prod_i \sigma^x_{i}$$. On a periodic chain, every $\sigma^z_i$ appears twice in the product on the left-hand side. Hence
$$U\left(\prod_{i}\tau^x_{i-\frac{1}{2}}\right)U^{-1} = \prod_i \sigma^x_i.$$
On the other hand,
$\prod_i\tau^z_{i-\frac{1}{2}}\sigma^x_i\tau^z_{i+\frac{1}{2}} = \prod_i\sigma^x_i$ and $\prod_i\tau^y_{i-\frac{1}{2}}\sigma^x_i\tau^y_{i+\frac{1}{2}} = \prod_i\sigma^x_i.$
Therefore, any map preserving the set of these two terms, including the possibility of exchanging them and/or translating their indices, implies
$$U\left(\prod_i\sigma_i^x\right)U^{-1} = \prod_i\sigma_i^x.$$
Combining the two conclusions would require $$\prod_i\sigma_i^x = \prod_i\tau^x_{i-\frac{1}{2}}$$
This is false as an operator identity on the full periodic Hilbert space. Hence no invertible operator can implement this Pauli duality. In particular, the duality cannot be realized by a local unitary transformation. 

This obstruction is analogous to the standard Kramers-Wannier duality on a periodic chain\cite{Seiberg_2024_2, Seifnashri_2024}: if the map $X_j\mapsto Z_{j-1}Z_j$ were implemented by a unitary operator, then $U\left(\prod_jX_j\right)U^{-1}=\prod_jZ_{j-1}Z_j=1$. Since $U$ is invertible, this would imply $\prod_jX_j=1$, which is false as an operator identity on the full Hilbert space.

\section{\label{sec:KW_Cliffordmap}The Pauli string mapping by KW transformation}
We now explain why a KW-type non-invertible duality can still include a mapping of the Pauli spectrum within the projected symmetry sector.
Considering a Hamiltonian satisfying $H(h)\mathbf{D}=\mathbf{D}H(1-h),$ where the KW operator\cite{Seiberg_2024_1, Seiberg_2024_2, Parayil_Mana_2024} can be decomposed as $\mathbf{D}=\mathbf{\tilde{D}}\mathbf{\eta}.$
Here $\mathbf{\tilde{D}}$ is a unitary Clifford operator, while $\mathbf{\eta}$ is the projector onto the even parity sector, $\mathbf{\eta}=\frac{1+P}{2}$.
Because of the projector, $\mathbf{D}$ is non-invertible and should not be regarded as a unitary Clifford circuit. 

Let $\ket{\Psi_+}$ be an even parity eigenstate of $H(1-h)$,
$$H(1-h)\ket{\Psi_+}=E\ket{\Psi_+},\qquad\eta\ket{\Psi_+} = \ket{\Psi_+}.$$
Then
$$H(h)\mathbf{D}\ket{\Psi_+}=\mathbf{D}H(1-h)\ket{\Psi_+}=E\mathbf{D}\ket{\Psi_+}.$$
Therefore,
$\ket{\Psi_+'}=\mathbf{D}\ket{\Psi_+}$
is an eigenstate of $H(h)$.

For a Pauli string $O$, we then have 
\begin{align*}
    \bra{\Psi'_+}O\ket{\Psi'_+}&=\bra{\Psi_+}\mathbf{D}^{\dagger}O\mathbf{D}\ket{\Psi_+}\\
    &= \bra{\Psi_+}\mathbf{\eta}\mathbf{\tilde{D}}^{\dagger}O\mathbf{\tilde{D}}\mathbf{\eta}\ket{\Psi_+}\\
    &= \bra{\Psi_+}\mathbf{\tilde{D}}^{\dagger}O\mathbf{\tilde{D}}\ket{\Psi_+}.
\end{align*}
Because $\mathbf{\tilde{D}}$ is Clifford, $\mathbf{\tilde{D}}^{\dagger}O\mathbf{\tilde{D}}$ is another Pauli string. Hence, within the even parity sector, the KW duality maps the Pauli spectrum of $\ket{\Psi_+}$ to that of the dual eigenstate $\ket{\Psi_+'}$ by relabeling Pauli strings.

A similar argument applies to the KT transformation. As discussed in Ref.\cite{Parayil_Mana_2024}, the KT transformation on a ring can be written as a product of KW-type transformations on the even and odd sublattices, schematically
$$\overline{KT}=\tilde{\mathbf D'}^{\,e}\eta^e\tilde{\mathbf D'}^{\,e}\eta^e.$$
Here $\eta^{e,o}$ are the projectors onto the corresponding even parity sectors, and $\tilde{\mathbf D'}^{\,e,o}$ are unitary Clifford parts. Therefore, for states in the projected even sectors on even and odd site, respectively, the same argument as above shows that KT transformation induces a relabeling of  Pauli strings.

\bibliography{apssamp.bib}

\end{document}